\documentclass{moriond}

\def\be{\begin{equation}}
\def\ee{\end{equation}}
\def\bea{\begin{eqnarray}}
\def\eea{\end{eqnarray}}

\def\ttbar{\ensuremath{\mathrm{t\bar{t}}}\xspace}
\def\mt{\ensuremath{{m_\mathrm{t}}}\xspace}
\def\stt{\ensuremath{\sigma_{\ttbar}}\xspace}
\def\sqrts{\ensuremath{\sqrt{\mathrm{s}}}\xspace}
\def\TeV{\ensuremath{\;\mathrm{TeV}}\xspace}
\def\GeV{\ensuremath{\;\mathrm{GeV}}\xspace}
\def\pt{\ensuremath{{p_\mathrm{t}}}\xspace}
\def\fbinv{\ensuremath{\;\mathrm{fb}^{-1}}\xspace}
\def\msbar{\ensuremath{\overline{\mathrm{MS}}}\xspace}

\usepackage{xspace}

\newcommand{\Photo}{}

\begin{document}
\vspace*{4cm}
\title{Top Quark Mass and Cross Sections in ATLAS and CMS}

\author{Matteo M. Defranchis - for the ATLAS and CMS Collaborations}

\address{Experimental Physics Department, CERN, Geneva (Switzerland)}

\maketitle\abstracts{
With the large data set delivered during the second run of the CERN LHC, inclusive and differential measurements of the top quark-antiquark production cross section at the ATLAS and CMS experiments often reach a precision that is  comparable or superior to that of the corresponding theoretical calculations.  The results of these measurements can be used to test the validity of perturbative quantum chromodynamics (QCD), and to precisely extract the values of parameters of the QCD Lagrangian such as the top quark mass (\mt).  The value of \mt can also be measured, with higher precision,  by fully or partially reconstructing the invariant mass of the top quark decay products. However, the results of these measurements lack a clear theoretical interpretation. In these proceedings,  the most recent measurements of the top quark mass and cross sections by ATLAS and CMS are presented.}

\section{Introduction}

At the CERN LHC, top quarks are mainly produced in quark-antiquark (\ttbar) pairs via the mechanism of gluon fusion. The \ttbar production cross section (\stt) can be calculated in quantum chromodynamics (QCD), either inclusively or differentially, using a perturbative approach. The inclusive \stt is currently known at the next-to-next-to-leading order (NNLO) in perturbation theory, including next-to-next-to-leading logarithmic (NNLL) corrections, with a precision of about 5\%~\cite{Czakon:2011xx}.  The ATLAS~\cite{Aad:2008zzm} and CMS~\cite{Chatrchyan:2008aa} experiments have performed measurements of the inclusive \stt at different centre-of-mass energies, often outperforming the precision of the theoretical calculations. In this contribution, two new measurements~\cite{ATLAS:2021xhc,CMS:2021jig} of \stt at $\sqrts = 5.02\TeV$ by the ATLAS and CMS Collaborations, respectively, are presented for the first time.

With the large data set delivered by the LHC during Run~2, differential and multi-differential measurements have become increasingly more precise in the less populated phase space, including the one with highly boosted top quarks. Differential measurements are particularly sensitive to the details of the parton distribution functions (PDFs) of the proton, and can be used to test the validity of theoretical calculations and Monte Carlo (MC) predictions, or to extract the values of QCD parameters such as \mt or the strong coupling constant~\cite{Sirunyan:2019zvx}.  In this contribution, a new measurement~\cite{CMS:2021fhl} of single and double-differential \ttbar cross sections in the $\ell$+jets final state by the CMS Collaboration  is presented for the first time. This is also the first analysis in which the resolved and boosted distributions are unfolded simultaneously. Significant progress has also been made on the theoretical side, as single- and double-differential calculations are becoming available at NNLO for an increasing number of variables~\cite{Catani:2019hip}.

The value of \mt can also be measured directly by comparing MC predictions to variables related to the invariant mass of the top quark decay products. Unlike indirect measurements, the direct ones lack a clear theoretical interpretation due to the modelling of non-perturbative effects in the MC simulation.  In this contribution, a direct measurement of \mt obtained by the CMS Collaboration using single top events~\cite{CMS:2021txo} is presented for the first time. The relation between the results of direct and indirect approaches can be investigated using measurements of the invariant mass distribution of large-area jets containing the decay products of a boosted top quark, such as the one in Ref.~\cite{Sirunyan:2019rfa}. In fact, the value of \mt can be extracted from such measurements both with a direct approach,  i.e\@. by comparing to distributions obtained with MC generator, and by comparing to calculations obtained in the framework of soft collinear effective field theories~\cite{Fleming:2007qr}. 

\section{Cross section measurements and extractions of the top quark mass}

\begin{figure}[htb]
\begin{minipage}{0.4\linewidth}
\centerline{\includegraphics[width=\linewidth]{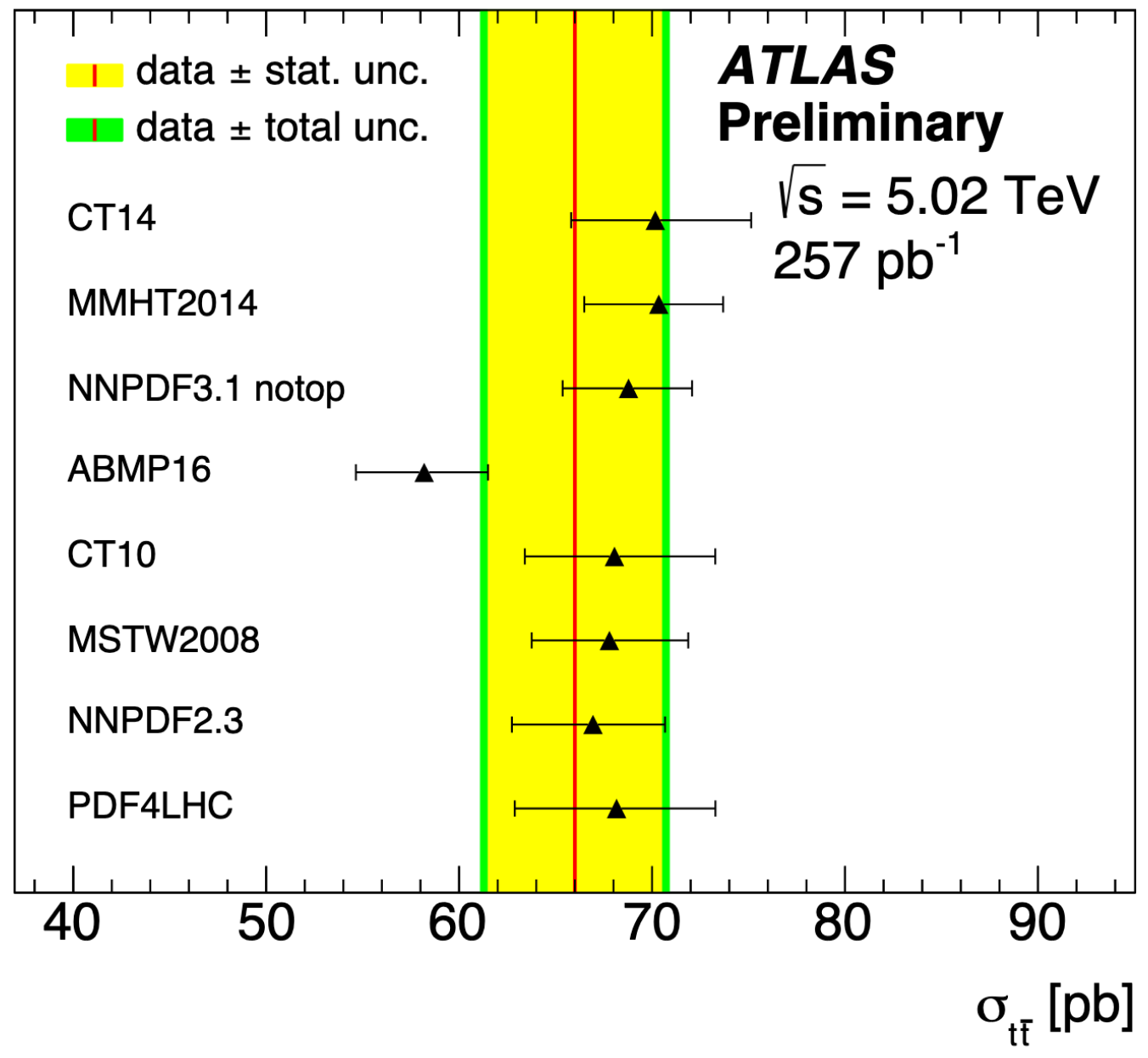}}
\end{minipage}
\hfill
\begin{minipage}{0.53\linewidth}
\centerline{\includegraphics[width=\linewidth]{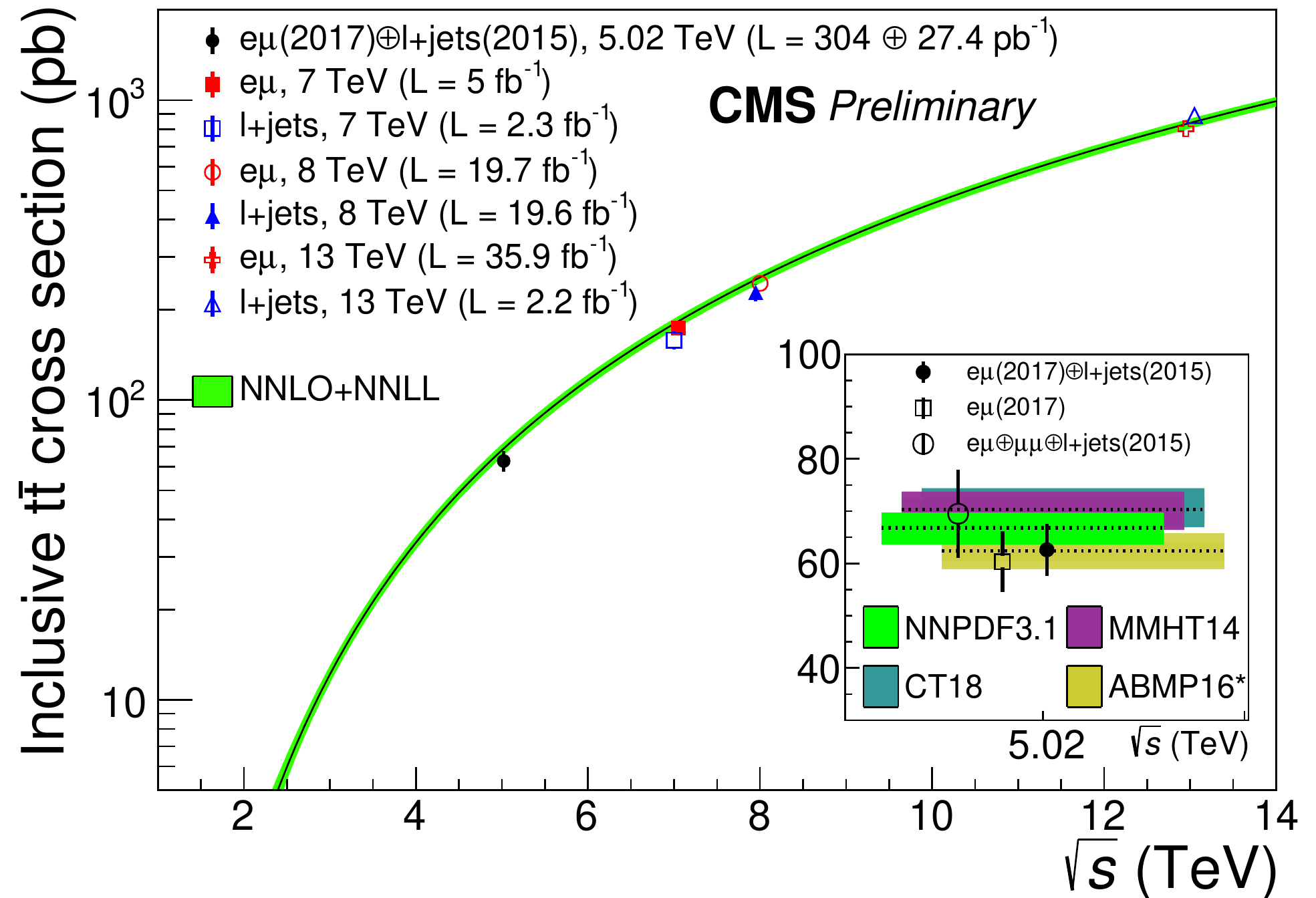}}
\end{minipage}
\caption{ATLAS~\protect\cite{ATLAS:2021xhc} (left) and CMS~\protect\cite{CMS:2021jig} (right) measurement of \stt at $\sqrts = 5.02 \TeV$ compared to NNLO+NNLL theoretical predictions~\protect\cite{Czakon:2011xx} with different sets of parton distribution functions.}
\label{fig:inclusive}
\end{figure}

In this contribution, two new measurements of the inclusive \stt at $\sqrts = 5.02\TeV$ by the ATLAS and CMS Collaborations, respectively, are presented~\cite{ATLAS:2021xhc,CMS:2021jig}. Both measurements are based on a small data set delivered by the LHC in 2017, corresponding to 257 and 304~$\mathrm{pb}^{-1}$ of proton-proton (pp) collisions for ATLAS and CMS, respectively. The ATLAS analysis consists of a maximum-likelihood fit in the dileptonic final states, while the CMS result is obtained using an event count method in the $\mathrm{e\mu}$ final state. The CMS result is then combined with a previous measurement obtained in the $\ell$+jets final state using a smaller data set collected in 2015~\cite{Sirunyan:2017ule}. The total uncertainty in the measured \stt is 7.5 (7.9)\% for ATLAS (CMS). The precision of both measurements is limited by the statistical uncertainty, and the results are in good agreement with the state-of-the-art theoretical predictions, as shown in Figure~\ref{fig:inclusive}.

To date, the most precise result of \stt at $\sqrts = 13\TeV$ was obtained by the ATLAS Collaboration in the $\mathrm{e\mu}$ final state using 36.1\fbinv of pp collision data~\cite{Aad:2019hzw} and has been used to determine the value of the top quark pole mass with a precision of about 2\GeV, limited by the QCD scale uncertainties. The total uncertainty in the measured \stt is 2.4\%, significantly smaller than the one of the NNLO+NNLL calculation~\cite{Czakon:2011xx}. Several other measurements at $\sqrts = 13\TeV$ have been performed by the ATLAS and CMS Collaborations in different decay channels~\cite{Aad:2020tmz,Sirunyan:2018goh}, including the one with with hadronically decaying $\tau$ leptons which is used by the CMS Collaboration to perform a test of the lepton universality in W boson decays~\cite{Sirunyan:2019guq}.

\begin{figure}[p]
\begin{minipage}{0.49\linewidth}
\centerline{\includegraphics[width=\linewidth]{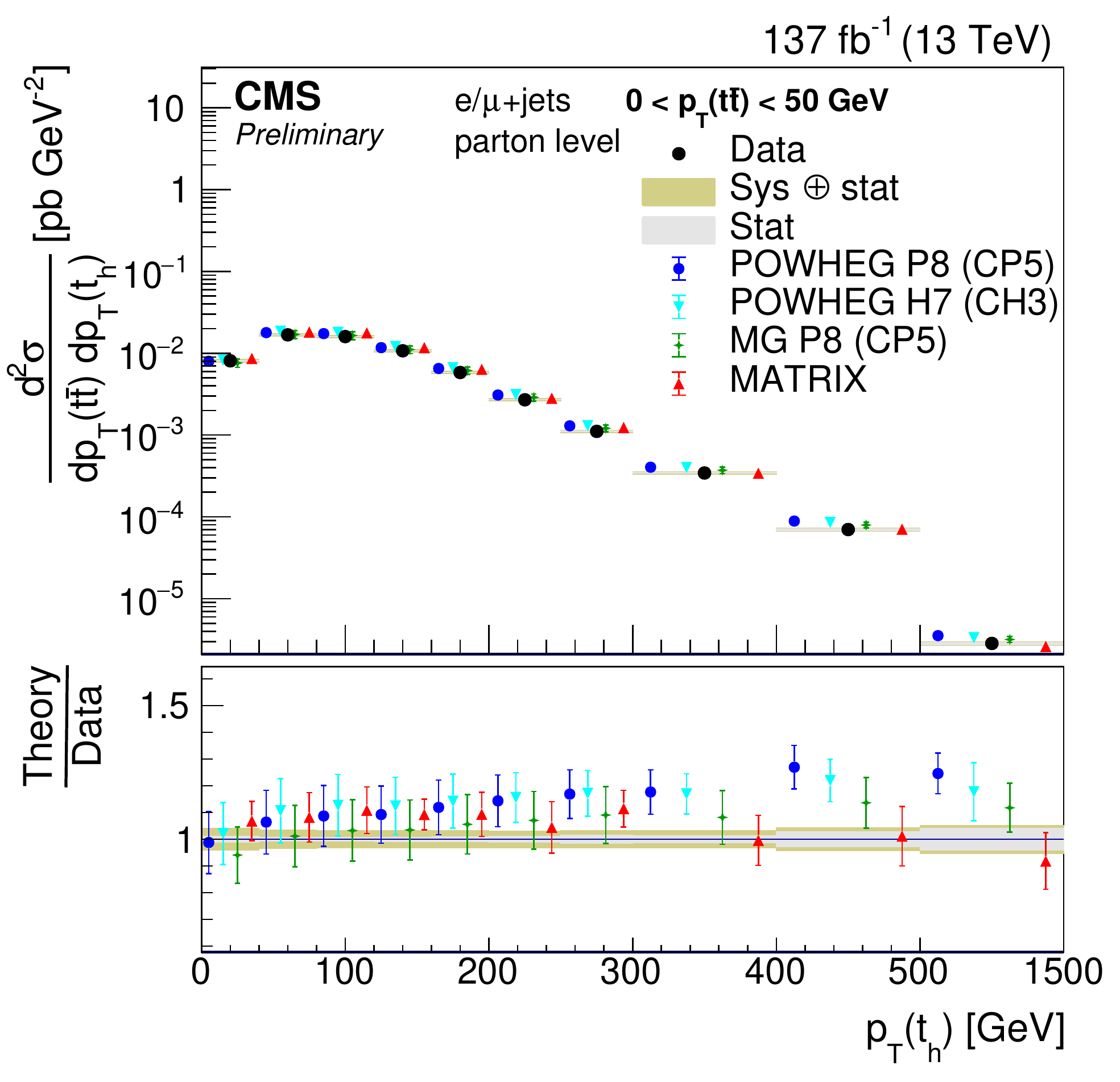}}
\end{minipage}
\hfill
\begin{minipage}{0.49\linewidth}
\centerline{\includegraphics[width=\linewidth]{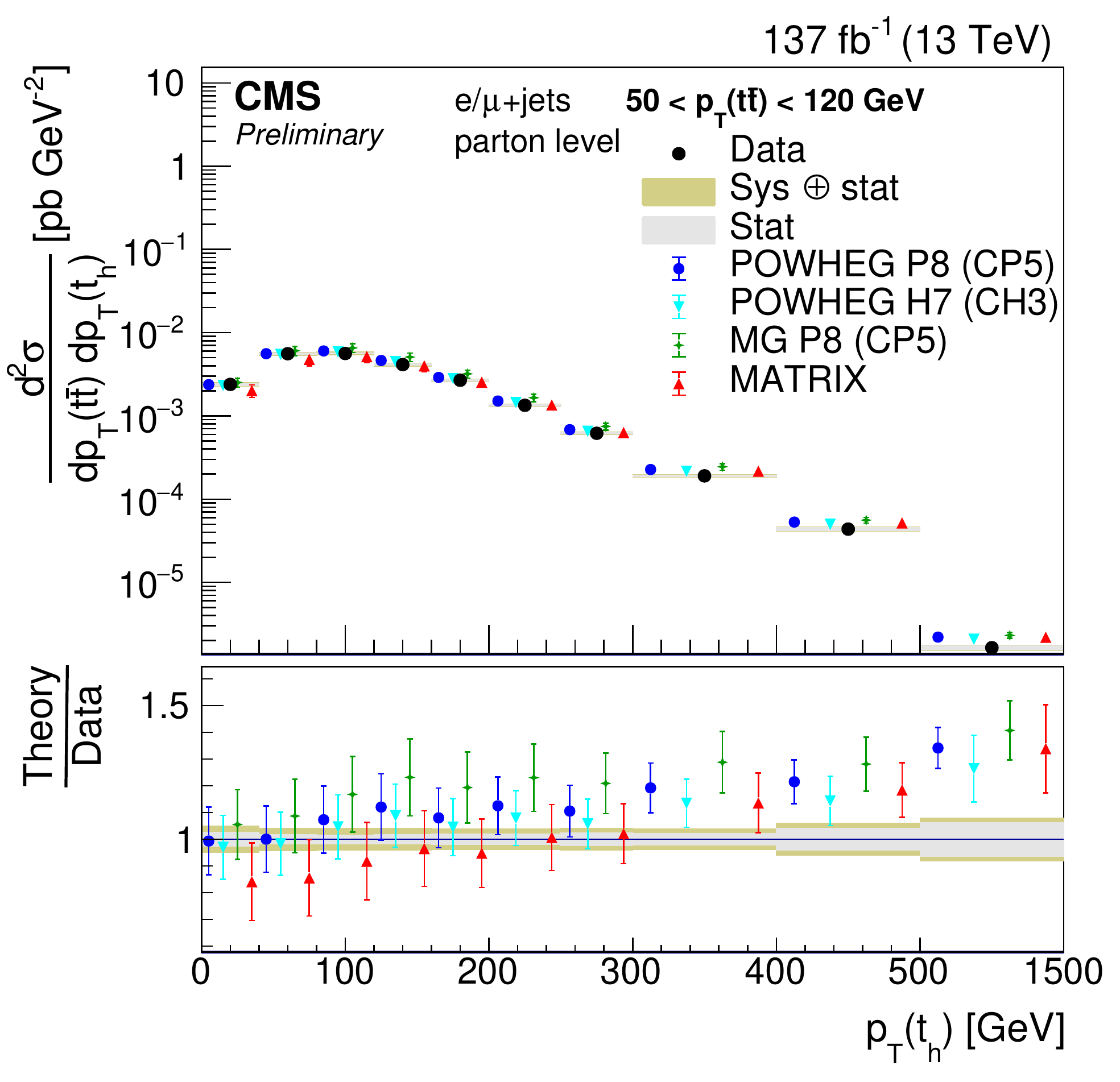}}
\end{minipage}

\begin{minipage}{0.49\linewidth}
\centerline{\includegraphics[width=\linewidth]{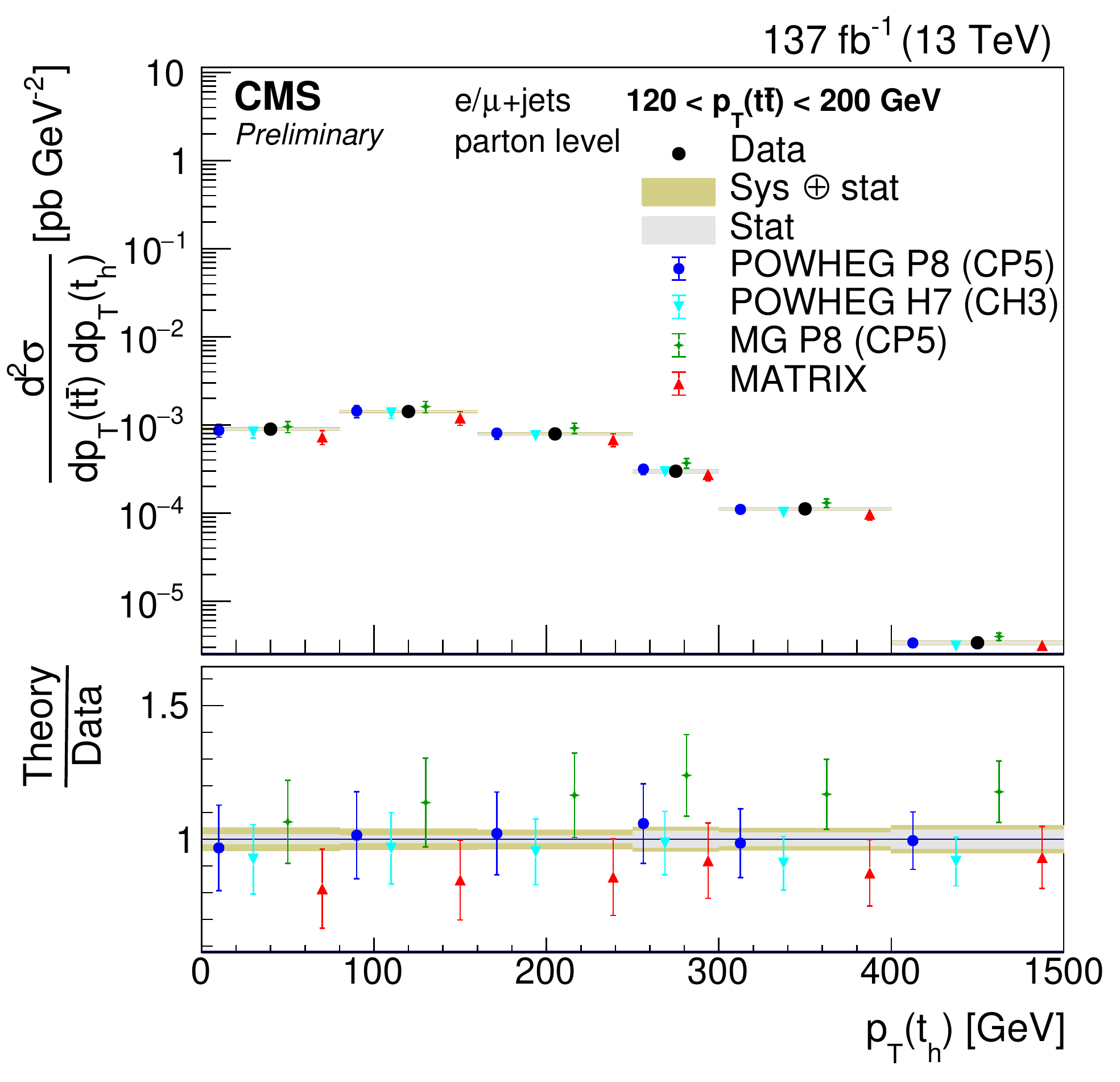}}
\end{minipage}
\hfill
\begin{minipage}{0.49\linewidth}
\centerline{\includegraphics[width=\linewidth]{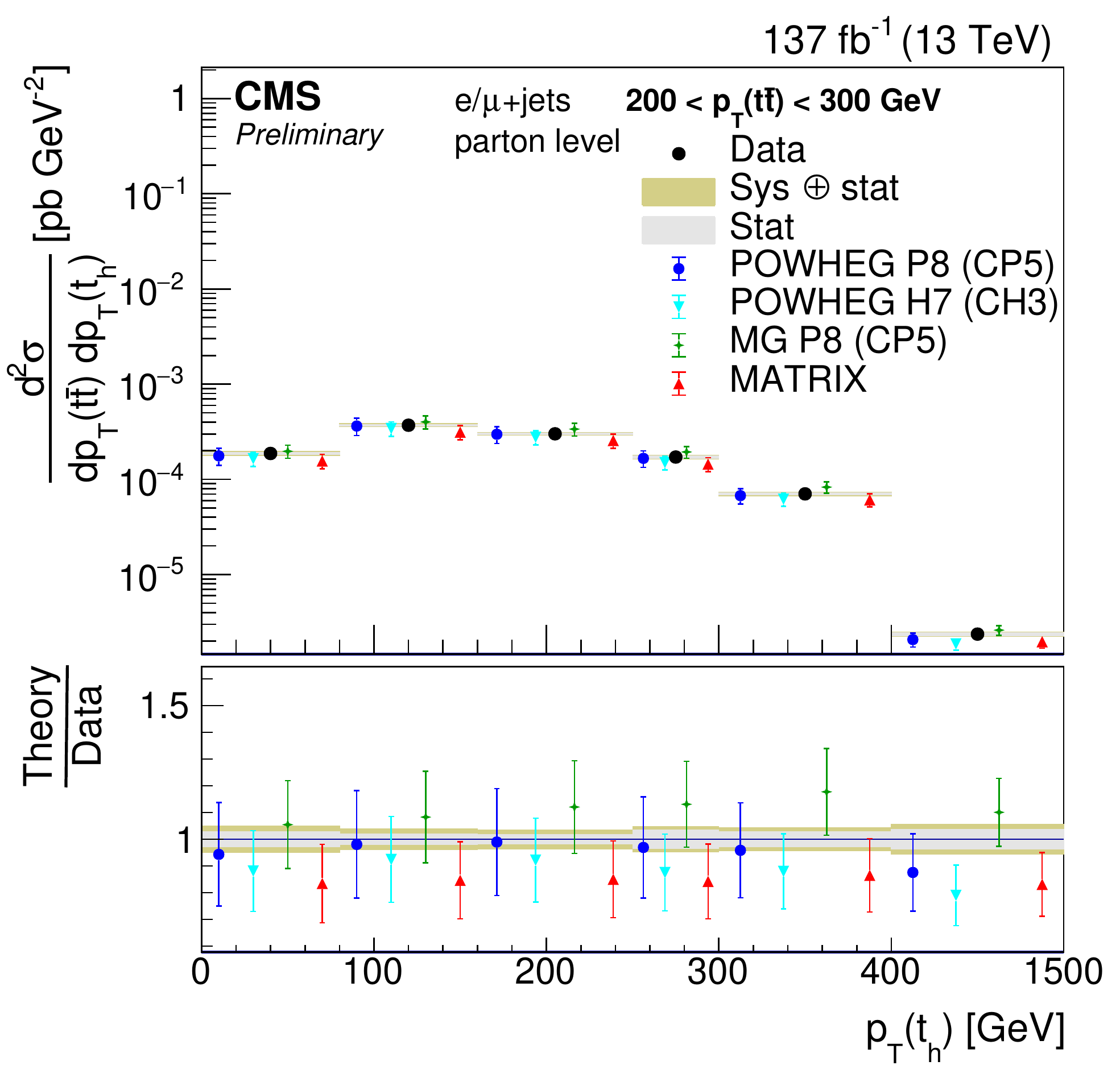}}
\end{minipage}

\begin{minipage}{0.49\linewidth}
\centerline{\includegraphics[width=\linewidth]{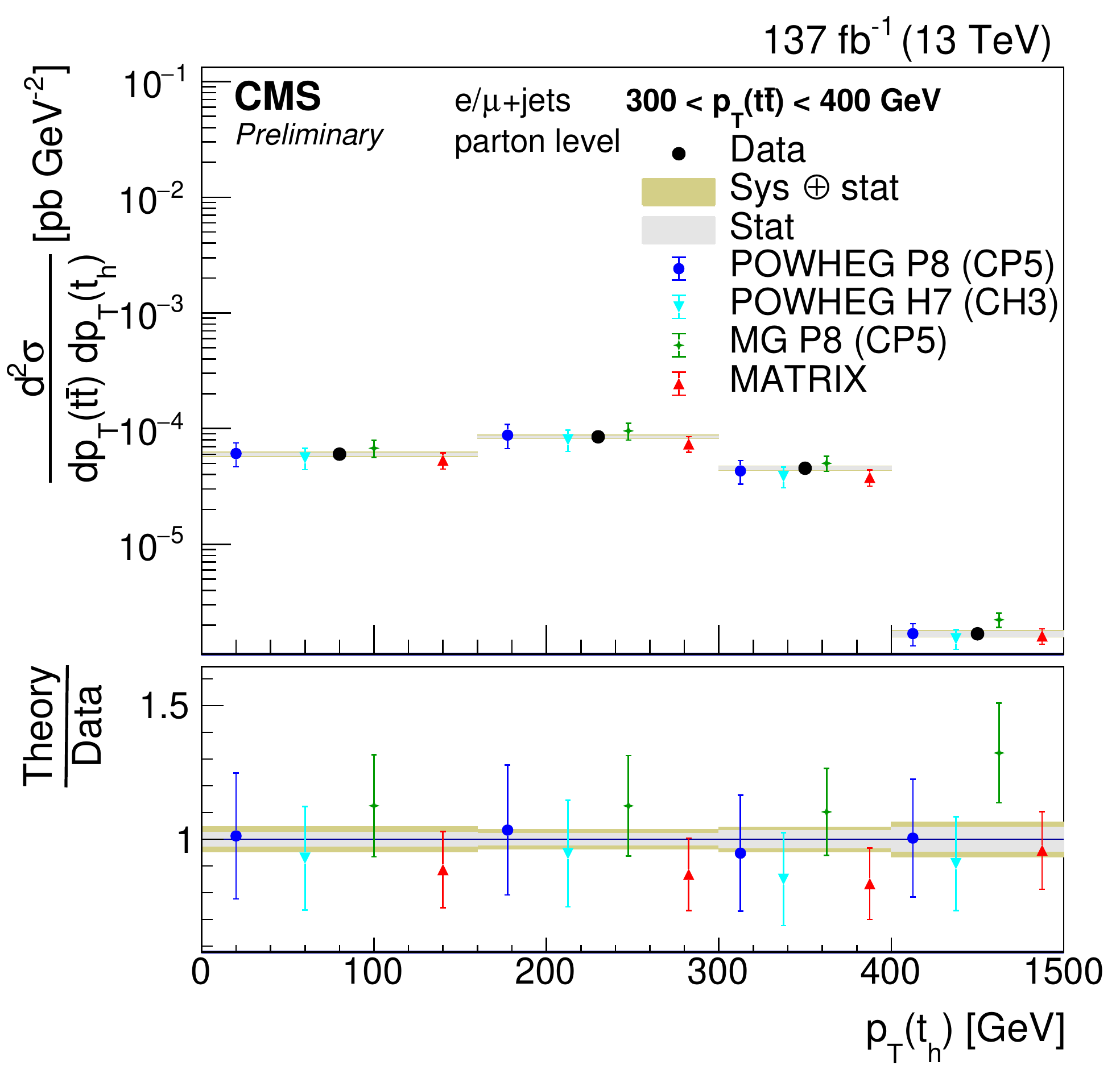}}
\end{minipage}
\hfill
\begin{minipage}{0.49\linewidth}
\centerline{\includegraphics[width=\linewidth]{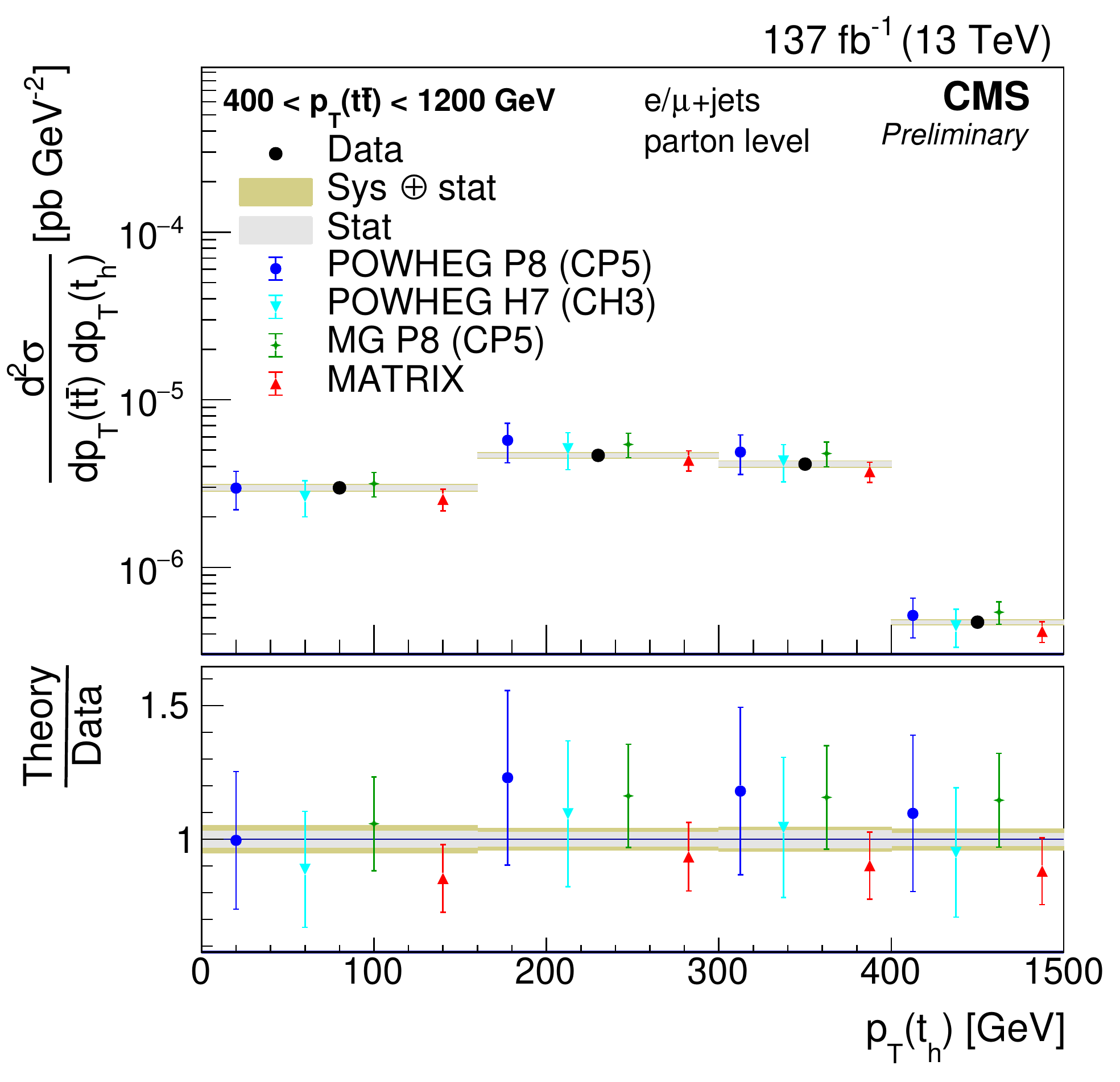}}
\end{minipage}

\caption{Measured normalised double-differential \ttbar cross section as a function of the \pt of the hadronically-decaying top quark and of the \pt of the \ttbar system, unfolded to the parton level in the full phase space \protect \cite{CMS:2021fhl}. The measurement is compared to state-of-the-art MC generators and NNLO theoretical predictions (\tt MATRIX~\protect\cite{Catani:2019hip}).}
\label{fig:diff}
\end{figure}

Single- and double-differential measurements of \stt at $\sqrts = 13\TeV$ are also performed by the two Collaborations in the various decay channels, both in the resolved and  boosted regimes. The measured cross sections are presented both absolute and normalized to the total cross section, and are unfolded to the particle and the parton level. Particle-level results are obtained in the fiducial phase space and can be used to test the predictions of different MC generators, while parton-level results are usually extrapolated to the full phase space and can be compared to fixed-order theoretical predictions. Furthermore, measurements of multi-differential cross sections can provide information about the correlation between different variables. A significant over-prediction of the spectrum of the transverse momentum (\pt) of the top quark in the boosted regime has been observed by both experiments in several measurements~\cite{Sirunyan:2019rfa,Sirunyan:2020vwa,Aad:2019ntk,Aaboud:2018eqg}, while double-differential measurements by the ATLAS Collaboration also highlight significant mismodelling of the \pt spectrum of the \ttbar system~\cite{Aad:2019ntk,Aad:2020nsf}.

In this contribution, a new measurement of single- and double-differential \ttbar cross sections at $\sqrts = 13\TeV$ by the CMS Collaboration~\cite{CMS:2021fhl} is presented for the first time. The analysis is performed in the $\ell$+jets final state using the full CMS Run~2 data set, corresponding to a total integrated luminosity of 137\fbinv. This is the first analysis in which the distributions in the resolved and boosted topologies are unfolded simultaneously. This is achieved by means of a minimum $\chi^2$ method with profiled systematic uncertainties, which also allows the total uncertainty in the measured cross sections to be significantly reduced. The inclusive \stt is also measured with a precision of 3.1\%, to date the most precise result in the $\ell$+jets channel. The measured \stt is found to be in good agreement with the state-of-the-art theoretical prediction, and the compatibility between the measured distributions and MC or fixed-order theoretical predictions is also quantitatively assessed. Similarly to previous ATLAS and CMS measurements, a softer-than-predicted spectrum of the top quark \pt is observed,  as shown, e.g\@.  in Figure~\ref{fig:diff}.

\begin{figure}[htb]
\centerline{\includegraphics[width=.6\linewidth]{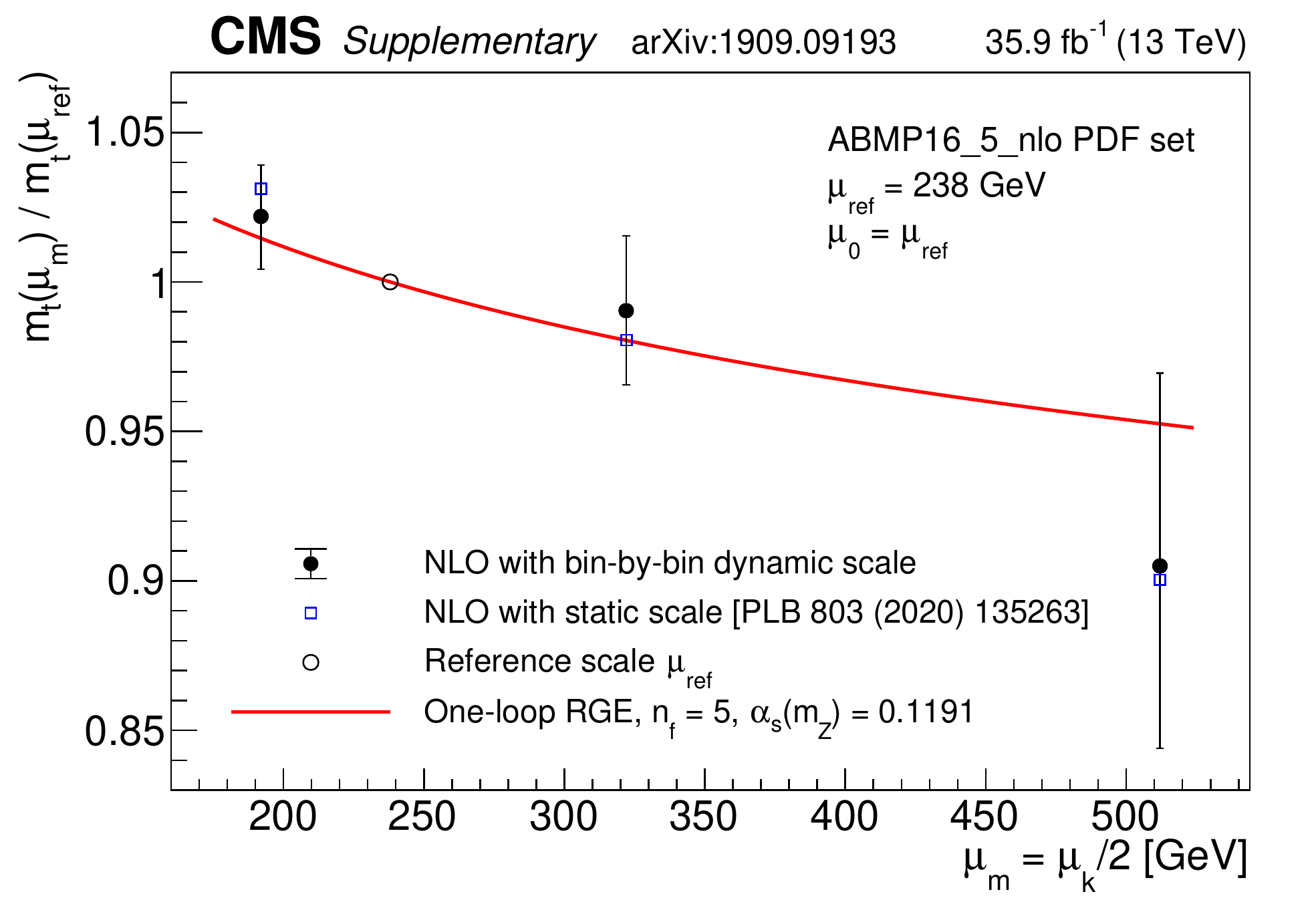}}
\caption{Measured scale dependence (running) of the value of top quark mass in the \msbar scheme as a function of $\mu_m = m_{\ttbar}/2$, obtained using NLO theoretical predictions with dynamic (full dots) or fixed (hollow squares) QCD scales \protect \cite{Sirunyan:2019jyn}. The result is in good agreement with the one-loop solution of the corresponding renormalisation group equation (line).}
\label{fig:running}
\end{figure}

Finally, a differential measurement of the \ttbar production cross section as a function of the invariant mass of the \ttbar system ($m_{\ttbar}$) is used by the CMS Collaboration to perform the first investigation of the running of the top quark mass~\cite{Sirunyan:2019jyn}, an effect analogous to the better-known running of the strong coupling constant.  The cross section is measured using 35.9\fbinv of pp collision data at $\sqrts = 13\TeV$ and is unfolded to the parton level by means of a maximum-likelihood method. The running is then extracted by comparing the measured cross section to NLO theoretical predictions in the modified minimal subtraction (\msbar) renormalisation scheme.  The result is found to be in good agreement with the one-loop solution of the corresponding renormalisation group equation, as shown in Figure~\ref{fig:running}. Furthermore, a hypothetical no-running scenario is excluded at above 95\% confidence level.

\section{Direct measurements of the top quark mass}

Direct measurements of \mt are performed by fully or partially reconstructing the invariant mass of the top quark decay products. In this contribution, the two most recent results by ATLAS and CMS are presented. Both analyses are performed in a different phase space compared to most direct measurements of \mt, and therefore provide additional information for future combinations of ATLAS and CMS results.

\begin{figure}[htb]
\begin{minipage}{0.49\linewidth}
\centerline{\includegraphics[width=\linewidth]{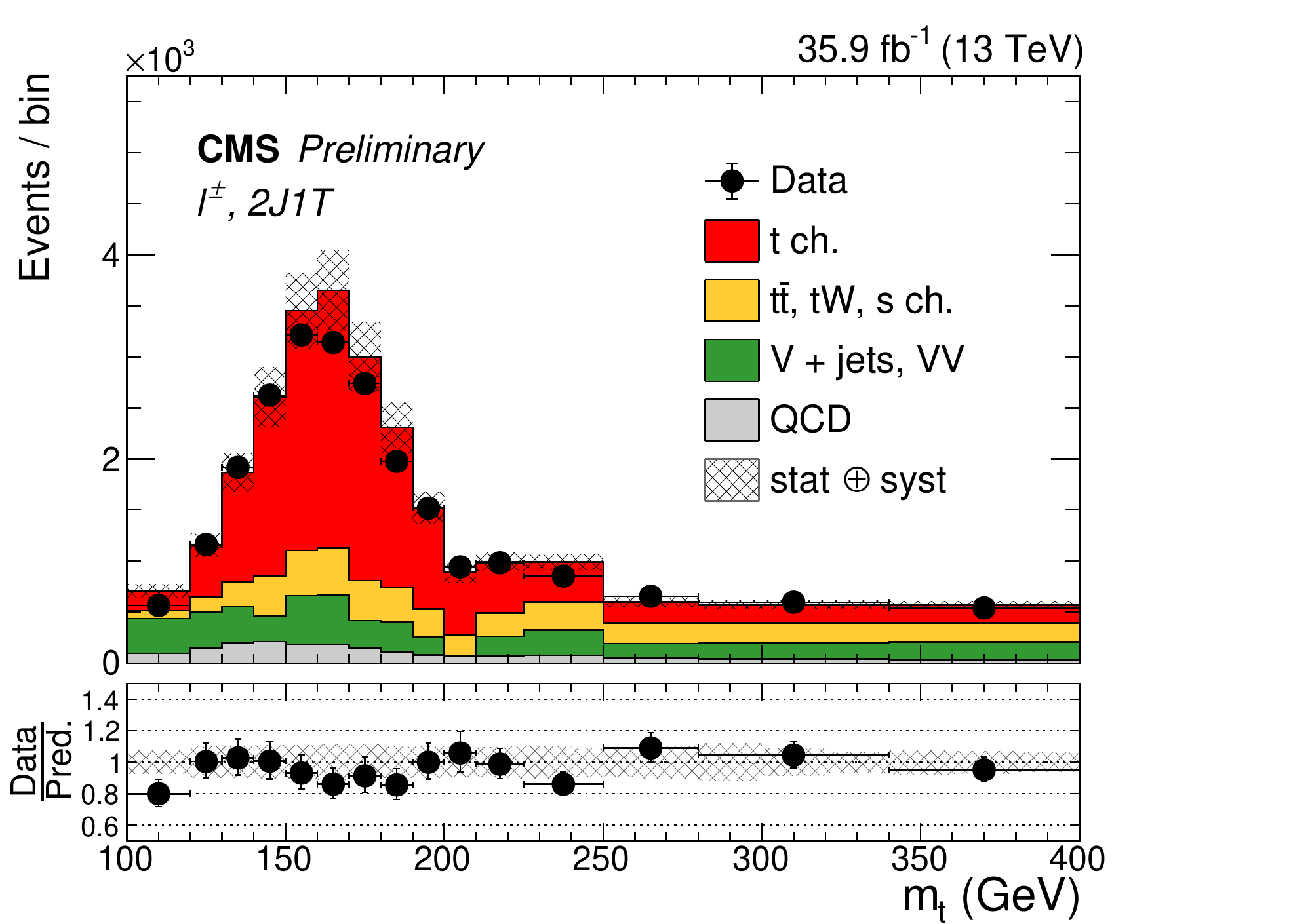}}
\end{minipage}
\hfill
\begin{minipage}{0.49\linewidth}
\centerline{\includegraphics[width=\linewidth]{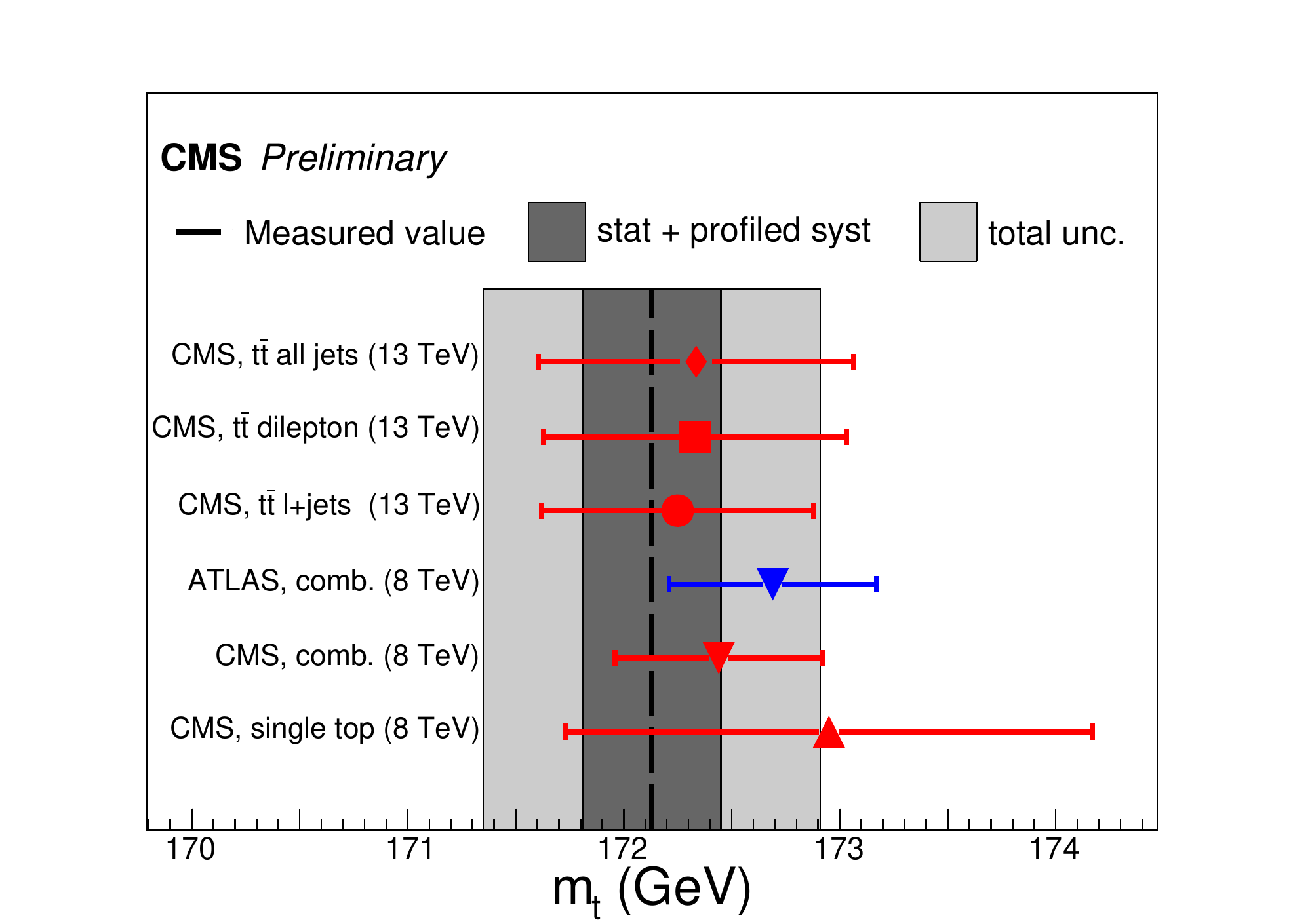}}
\end{minipage}
\caption{Reconstructed invariant mass of the top quark decay products in single top $t$-channel events (left),  and measured value of \mt compared to previous ATLAS and CMS results (right) \protect \cite{CMS:2021txo}.}
\label{fig:mt}
\end{figure}

The CMS measurement, which is presented here for the first time, consists of a functional fit to the reconstructed invariant mass of the top quark decay products in single top events in the $t$-channel (left-hand side of Figure~\ref{fig:mt}), and is performed using 35.9\fbinv of pp collision data~\cite{CMS:2021txo}. The top quark is required to decay leptonically, and the charge of the reconstructed lepton is used to distinguish between event containing top quarks and those containing top antiquarks. The value of \mt  is measured both inclusively, resulting in $\mt = 172.13 \pm 0.77\GeV$, and separately for top quarks and antiquarks.  The inclusive result is in good agreement with previous ATLAS and CMS measurements, as shown in Figure~\ref{fig:mt} (right), and is limited by the uncertainty in the jet energy scale calibration. The ratio and difference between the values of the top quark and antiquark masses are measured to be $R_{\mt} = m_\mathrm{\bar{t}} / \mt = 0.995 \pm 0.006$ and $\Delta \mt = \mt - m_\mathrm{\bar{t}} = 0.83 \,^{+0.77}_{-1.01} \GeV$, respectively, in good agreement with the hypothesis of conservation of the CPT symmetry.

The ATLAS measurement, performed with \ttbar candidate events in the $\ell$+jets final state,  is based on the reconstruction of the $m_{\ell\mu}$ variable, defined as the invariant mass between the lepton from the top quark decay and a soft, non-isolated muon originating from the leptonic decay of a B~hadron inside reconstructed a jet~\cite{ATLAS:2019ezb}. When the \ttbar system is correctly reconstructed, the jet containing the B~hadron originates from the b~quark from the leptonically-decaying top quark. This measurement, performed with 36.1\fbinv of pp collision data, is less sensitive to the calibration of the jet energy scale compared to conventional direct measurements, and therefore carries additional information when performing a combination of measurements. A dedicated algorithm was developed in the scope of this measurement to  identify the soft lepton originated from the B~hadron decay. Since this measurement is very sensitive to the details of the b~quark fragmentation, the parameters of the Bowler--Lund fragmentation function have been re-optimized by means of a fit to LEP and SLC data. In a dedicated study~\cite{ATLAS:2020iyi}, the parameters derived in the context of this analysis are found to well describe other distributions sensitive to the b~quark fragmentation. The top quark mass is measured to be $\mt = 178.48 \pm 0.40\, (\mathrm{stat}) \pm 0.67 \, (\mathrm{syst})\GeV$,  in agreement with the combination of previous ATLAS measurements~\cite{Aaboud:2018zbu} within 2.2 standard deviations. The precision of the measurement is limited by the uncertainty in the branching ratio of B~hadrons.

\section{Summary}

In this contribution, the most recent measurements of top quark mass and cross section at the ATLAS and CMS experiments at the CERN LHC are reviewed, including several new results. These include two measurements of the \ttbar production cross section in proton-proton collisions at $\sqrts = 5.02\TeV$ by the ATLAS and CMS Collaborations based on a small dataset delivered by the CERN LHC in 2017.  The results are in good agreement with state-of-the-art theoretical calculations, and are limited by the statistical uncertainty of the data. The CMS Collaboration has also published a measurement of single- and double-differential \ttbar cross sections in the $\ell$+jets final state based on the full LHC Run~2 dataset, where the distributions in the resolved and boosted topologies are unfolded simultaneously for the first time.  This is achieved by means of a profiled $\chi^2$ fit, which also allows the impact of systematic uncertainties to be significantly reduced. The \ttbar cross section is also measured inclusively with a total uncertainty of 3.1\%,  the most precise result in this channel, to date. In this analysis, a significantly softer-than-predicted spectrum of the top quark transverse momentum is observed, consistent with several previous measurements. Finally,  a direct measurement of the top quark mass using single top quark events in the $t$-channel is presented for the first time.   The value of the top quark mass is measured both inclusively and separately for top quarks and antiquarks, which allows for a test of the conservation of the CPT symmetry.

%If you more commonly use the method of square brackets in the line of text
%for citation than the superscript method,
%please note that you need  to adjust the punctuation
%so that the citation command appears after the punctuation mark.


\section*{References}
\begin{thebibliography}{99}


%\cite{Aad:2008zzm}
\bibitem{Aad:2008zzm}
ATLAS Collaboration,
%``The ATLAS Experiment at the CERN Large Hadron Collider,''
JINST \textbf{3} (2008) S08003
%doi:10.1088/1748-0221/3/08/S08003
%9337 citations counted in INSPIRE as of 02 May 2021

%\cite{Chatrchyan:2008aa}
\bibitem{Chatrchyan:2008aa}
CMS Collaboration,
%``The CMS Experiment at the CERN LHC,''
JINST \textbf{3} (2008) S08004
%doi:10.1088/1748-0221/3/08/S08004
%7708 citations counted in INSPIRE as of 02 May 2021

%\cite{Czakon:2011xx}
\bibitem{Czakon:2011xx}
M.~Czakon and A.~Mitov,
%``Top++: A Program for the Calculation of the Top-Pair Cross-Section at Hadron Colliders,''
Comput. Phys. Commun. \textbf{185} (2014) 2930
%doi:10.1016/j.cpc.2014.06.021
%[arXiv:1112.5675 [hep-ph]].
%1460 citations counted in INSPIRE as of 03 May 2021

%\cite{ATLAS:2021xhc}
\bibitem{ATLAS:2021xhc}
 ATLAS Collaboration,
%``Measurement of the $t\bar{t}$ production cross-section using dilepton events in $pp$ collisions at $\sqrt{s}=5.02$ TeV with the ATLAS detector,''
ATLAS-CONF-2021-003,
https://cds.cern.ch/record/2754223
%0 citations counted in INSPIRE as of 03 May 2021

%\cite{CMS:2021jig}
\bibitem{CMS:2021jig}
CMS Collaboration,
%``Measurement of the inclusive $\mathrm{t}\bar{\mathrm{t}}$ production cross section in pp collisions at $\sqrt{s} = 5.02\,\mathrm{TeV}$,''
CMS-PAS-TOP-20-004,
http://cds.cern.ch/record/2758333
%0 citations counted in INSPIRE as of 03 May 2021

%\cite{Sirunyan:2019zvx}
\bibitem{Sirunyan:2019zvx}
CMS Collaboration,
%``Measurement of $\mathrm{t\bar t}$ normalised multi-differential cross sections in pp collisions at $\sqrt s=13$ TeV, and simultaneous determination of the strong coupling strength, top quark pole mass, and parton distribution functions,''
Eur. Phys. J. C \textbf{80} (2020) 658
%doi:10.1140/epjc/s10052-020-7917-7
%[arXiv:1904.05237 [hep-ex]].
%63 citations counted in INSPIRE as of 03 May 2021

%\cite{CMS:2021fhl}
\bibitem{CMS:2021fhl}
CMS Collaboration,
%``Measurement of differential $\mathrm{t}\bar{\mathrm{t}}$ production cross sections in the full kinematic range using lepton+jets events from pp collisions at $\sqrt{s}=13~\mathrm{TeV}$,''
CMS-PAS-TOP-20-001,
https://cds.cern.ch/record/2759302
%0 citations counted in INSPIRE as of 03 May 2021

%\cite{Catani:2019hip}
\bibitem{Catani:2019hip}
S.~Catani, S.~Devoto, M.~Grazzini, S.~Kallweit and J.~Mazzitelli,
%``Top-quark pair production at the LHC: Fully differential QCD predictions at NNLO,''
JHEP \textbf{07} (2019) 100
%doi:10.1007/JHEP07(2019)100
%[arXiv:1906.06535 [hep-ph]].
%48 citations counted in INSPIRE as of 03 May 2021

%\cite{CMS:2021txo}
\bibitem{CMS:2021txo}
CMS Collaboration,
%``Measurement of the top quark mass in events with a single reconstructed top quark at $\sqrt{s}=13\, \mathrm{TeV}$,''
CMS-PAS-TOP-19-009,
http://cds.cern.ch/record/2759301
%0 citations counted in INSPIRE as of 03 May 2021

%\cite{Sirunyan:2019rfa}
\bibitem{Sirunyan:2019rfa}
CMS Collaboration,
%``Measurement of the Jet Mass Distribution and Top Quark Mass in Hadronic Decays of Boosted Top Quarks in $pp$ Collisions at $\sqrt{s} =$  TeV,''
Phys. Rev. Lett. \textbf{124} (2020) 202001
%doi:10.1103/PhysRevLett.124.202001
%[arXiv:1911.03800 [hep-ex]].
%10 citations counted in INSPIRE as of 03 May 2021

%\cite{Fleming:2007qr}
\bibitem{Fleming:2007qr}
S.~Fleming, A.~H.~Hoang, S.~Mantry and I.~W.~Stewart,
%``Jets from massive unstable particles: Top-mass determination,''
Phys. Rev. D \textbf{77} (2008) 074010
%doi:10.1103/PhysRevD.77.074010
%[arXiv:hep-ph/0703207 [hep-ph]].
%242 citations counted in INSPIRE as of 03 May 2021



%\cite{Sirunyan:2017ule}
\bibitem{Sirunyan:2017ule}
CMS Collaboration,
%``Measurement of the inclusive $ \mathrm{t}\overline{\mathrm{t}} $ cross section in pp collisions at $ \sqrt{s}=5.02 $ TeV using final states with at least one charged lepton,''
JHEP \textbf{03} (2018) 115
%doi:10.1007/JHEP03(2018)115
%[arXiv:1711.03143 [hep-ex]].
%35 citations counted in INSPIRE as of 04 May 2021

%\cite{Aad:2019hzw}
\bibitem{Aad:2019hzw}
ATLAS Collaboration,
%``Measurement of the $t\bar{t}$ production cross-section and lepton differential distributions in $e\mu $ dilepton events from $pp$ collisions at $\sqrt{s}=13\,\text {TeV}$ with the ATLAS detector,''
Eur. Phys. J. C \textbf{80} (2020) 528
%doi:10.1140/epjc/s10052-020-7907-9
%[arXiv:1910.08819 [hep-ex]].
%32 citations counted in INSPIRE as of 03 May 2021

%\cite{Aad:2020tmz}
\bibitem{Aad:2020tmz}
ATLAS Collaboration,
%``Measurement of the $t\bar{t}$ production cross-section in the lepton+jets channel at $\sqrt{s}=13$ TeV with the ATLAS experiment,''
Phys. Lett. B \textbf{810} (2020) 135797
%doi:10.1016/j.physletb.2020.135797
%[arXiv:2006.13076 [hep-ex]].
%10 citations counted in INSPIRE as of 03 May 2021

%\cite{Sirunyan:2018goh}
\bibitem{Sirunyan:2018goh}
CMS Collaboration,
%``Measurement of the $\mathrm{t}\overline{\mathrm{t}}$ production cross section, the top quark mass, and the strong coupling constant using dilepton events in pp collisions at $\sqrt{s} =$ 13 TeV,''
Eur. Phys. J. C \textbf{79} (2019) 368
%doi:10.1140/epjc/s10052-019-6863-8
%[arXiv:1812.10505 [hep-ex]].
%80 citations counted in INSPIRE as of 03 May 2021

%\cite{Sirunyan:2019guq}
\bibitem{Sirunyan:2019guq}
CMS Collaboration,
%``Measurement of the top quark pair production cross section in dilepton final states containing one $\tau$ lepton in pp collisions at $\sqrt{s}=$ 13 TeV,''
JHEP \textbf{02} (2020) 191
%doi:10.1007/JHEP02(2020)191
%[arXiv:1911.13204 [hep-ex]].
%7 citations counted in INSPIRE as of 03 May 2021





%\cite{Sirunyan:2020vwa}
\bibitem{Sirunyan:2020vwa}
CMS Collaboration,
%``Measurement of differential $\mathrm{t\bar{t}}$ production cross sections using top quarks at large transverse momenta in $pp$ collisions at $\sqrt{s} =$ 13 TeV,''
Phys. Rev. D \textbf{103} (2021) 052008
%doi:10.1103/PhysRevD.103.052008
%[arXiv:2008.07860 [hep-ex]].
%5 citations counted in INSPIRE as of 03 May 2021



%\cite{Aad:2019ntk}
\bibitem{Aad:2019ntk}
ATLAS Collaboration,
%``Measurements of top-quark pair differential and double-differential cross-sections in the $\ell$+jets channel with $pp$ collisions at $\sqrt{s}=13$ TeV using the ATLAS detector,''
Eur. Phys. J. C \textbf{79} (2019) 1028
[erratum: Eur. Phys. J. C \textbf{80} (2020) 1092]
%doi:10.1140/epjc/s10052-019-7525-6
%[arXiv:1908.07305 [hep-ex]].
%36 citations counted in INSPIRE as of 03 May 2021

%\cite{Aaboud:2018eqg}
\bibitem{Aaboud:2018eqg}
ATLAS Collaboration,
%``Measurements of $t\bar{t}$ differential cross-sections of highly boosted top quarks decaying to all-hadronic final states in $pp$ collisions at $\sqrt{s}=13\,$ TeV using the ATLAS detector,''
Phys. Rev. D \textbf{98} (2018) 012003
%doi:10.1103/PhysRevD.98.012003
%[arXiv:1801.02052 [hep-ex]].
%80 citations counted in INSPIRE as of 03 May 2021


%\cite{Aad:2020nsf}
\bibitem{Aad:2020nsf}
ATLAS Collaboration,
%``Measurements of top-quark pair single- and double-differential cross-sections in the all-hadronic channel in $pp$ collisions at $\sqrt{s}=13~\textrm{TeV}$ using the ATLAS detector,''
JHEP \textbf{01} (2021) 033
%doi:10.1007/JHEP01(2021)033
%[arXiv:2006.09274 [hep-ex]].
%10 citations counted in INSPIRE as of 03 May 2021

%\cite{Sirunyan:2019jyn}
\bibitem{Sirunyan:2019jyn}
CMS Collaboration,
%``Running of the top quark mass from proton-proton collisions at $\sqrt{s} =$ 13TeV,''
Phys. Lett. B \textbf{803} (2020) 135263
%doi:10.1016/j.physletb.2020.135263
%[arXiv:1909.09193 [hep-ex]].
%22 citations counted in INSPIRE as of 03 May 2021


%\cite{ATLAS:2019ezb}
\bibitem{ATLAS:2019ezb}
ATLAS Collaboration,
%``Measurement of the top quark mass using a leptonic invariant mass in pp collisions at $sqrt{s}$ = 13 TeV with the ATLAS detector,''
ATLAS-CONF-2019-046,
https://cds.cern.ch/record/2693954
%5 citations counted in INSPIRE as of 03 May 2021

%\cite{ATLAS:2020iyi}
\bibitem{ATLAS:2020iyi}
ATLAS Collaboration,
%``Measurements of $b$-jet moments sensitive to $b$-quark fragmentation in $t \bar{t}$ events at the LHC with the ATLAS detector,''
ATLAS-CONF-2020-050,
https://cds.cern.ch/record/2730444
%0 citations counted in INSPIRE as of 03 May 2021


%\cite{Aaboud:2018zbu}
\bibitem{Aaboud:2018zbu}
ATLAS Collaboration,
%``Measurement of the top quark mass in the $t\bar{t}\rightarrow $ lepton+jets channel from $\sqrt{s}=8$  TeV ATLAS data and combination with previous results,''
Eur. Phys. J. C \textbf{79} (2019) 290
%doi:10.1140/epjc/s10052-019-6757-9
%[arXiv:1810.01772 [hep-ex]].
%71 citations counted in INSPIRE as of 04 May 2021

\end{thebibliography}
\end{document}